\documentclass[12pt]{article}

\usepackage{amsmath}
\usepackage{cite}

\numberwithin{equation}{section}

\title{Unfolded equations for massive higher spin
supermultiplets in $AdS_3$}

\author{I.L. Buchbinder${}^{ab}$\thanks{joseph@tspu.edu.ru},
T.V. Snegirev${}^{ac}$\thanks{snegirev@tspu.edu.ru}, Yu.M.
Zinoviev$^d$\thanks{Yurii.Zinoviev@ihep.ru}
\\[0.5cm]
\it ${}^a$Department of Theoretical Physics,\\
\it Tomsk State Pedagogical University,\\
\it Tomsk 634061, Russia\\[0.3cm]
\it ${}^b$National Research Tomsk State University, Russia\\[0.3cm]
\it ${}^c$National Research Tomsk Polytechnic University,\\
Tomsk 634050, Russia\\[0.3cm]
\it ${}^d$Institute for High Energy Physics,\\
\it of National Research Center "Kurchatov Institute" \\
\it Protvino, Moscow Region, 142280, Russia}

\date{}

\begin{document}

\maketitle

\begin{abstract}
In this paper we give an explicit construction of unfolded equations
for massive higher spin supermultiplets of the minimal
$(1,0)$ supersymmetry in $AdS_3$ space. For that purpose we use an
unfolded formulation for massive bosonic and fermionic higher spins
and find supertransformations leaving appropriate set of unfolded
equations invariant. We provide two general supermultiplets $(s,
s+1/2)$ and $(s, s-1/2)$ with arbitrary integer s, as well as a
number of lower spin examples.
\end{abstract}

\thispagestyle{empty}
\newpage
\setcounter{page}{1}

\section*{Introduction}

Among all the higher spin symmetries a supersymmetry still plays a
distinguished role. It is enough to recall that all massive states in
the superstring theory are nicely organized into massive higher spin
supermultiplets. The classification of massless and massive
supermultiplets (depending on the space-time dimensions, number of
supersymmetries and type of fermions) is purely algebraic task and is
very well understood now. But as far as the concrete realization in
terms of Lagrangians and/or equations of motion is concerned there is
a striking difference between massless and massive cases. For the
massless supermultiplets the realization can be straightforwardly
constructed (both in components as well as in superfields) simply
because the supertransformations always have the same simple pattern:
$$
\delta B \sim F \eta, \qquad \delta F \sim \partial B \eta
$$
where $B$ and $F$ are bosonic and fermionic fields and $\eta$ ---
parameter of the supertransformations. But switching from the massless
to the massive case one has to introduce a lot of complicated higher
derivatives corrections to the supertransformations without any
evident pattern and the higher the spin of fields entering the
supermultiplet the higher the number of derivatives one has to
consider.

In the four-dimensional Minkowski space the solution in components
was proposed in \cite{Zin07a} based on the gauge invariant
description of massive higher spin bosonic \cite{Zin01} and
fermionic \cite{Met06} fields\footnote{Construction of the
Lagrangian superfield description for massive superspins 1 and 3/2
has been initiated by the papers \cite{BGLP02} and \cite{BGPL02}
(see also \cite{OS}, \cite{GK}, \cite{BGKP}, \cite{GK_1},
\cite{GKT}, \cite{GKo}) while analogous description for massive
supermultiplets with arbitrary superspin is absent up to now.
Lagrangian description for massless higher superspin theories is
developed much better \cite{KS}, \cite{GK1}, \cite {GKo1},
\cite{BKo}.}. The main idea was that the massive supermultiplet can
be constructed out of the appropriate set of massless ones in the
same way as the gauge invariant description of massive higher spin
particles can be constructed using an appropriate set of massless
ones. In spite of the large number of fields involved such
construction appears to be pretty straightforward. At the same time
the meaning of these complicated corrections to the
supertransformations that one has to introduce working with
non-gauge invariant description of massive particles becomes clear.
Namely, they are just the restoring gauge transformations that
appear when one tries to exclude all Stueckelberg fields fixing all
gauge symmetries.

Recently, using the same approach, we have constructed an explicit
Lagrangian formulation for massive higher spin supermultiplets in
three-dimensional Minkowski space \cite{BSZ15} based on our previous
works on the gauge invariant formulation for massive bosonic
\cite{BSZ12a} and fermionic \cite{BSZ14a} higher spins in three
dimensions\footnote{Superfield approach to three dimensional higher
spin supersymmetric models proposed in \cite{KO}.}. The aim of the
current work is to investigate massive higher spin supermultiplets
in the three-dimensional anti de Sitter space. Note that the
approach under consideration leads to on-shell supersymmetric models
where the auxiliary fields allowing to close the supersymmetry
algebra are absent. As is well known \cite{AT86}, $AdS_3$ space is
special because all $AdS_3$ superalgebras (as well as $AdS_3$
algebra itself) factorize into "left" and "right" parts. For the
case of simplest (1,0) superalgebra (the one we are working here
with) it has the form:
$$
OSp(1,2) \otimes Sp(2)
$$
so that we have supersymmetry in the "left" sector only. It means
that the minimal massive supermultiplet must contain just one
bosonic and one fermionic degrees of freedom. To realize such
supermultiplets we will use a so called unfolded formalism
\cite{Vas91a,Vas94} which apparently is a most simple and efficient
way\footnote{In principle the description of such supermultiplets can
be done analogous to \cite{BSZ15}, \cite{BSZ12a}, \cite{BSZ14a}.
However the consideration turns out to be more complicated.}. For the
dimensions $d \ge 4$ complete unfolded description of massive bosonic
higher spins has been constructed in \cite{PV10}. In three dimensions
the authors of \cite{BPSS15} suggested unfolded equations for the
infinite chain of gauge invariant zero-forms and showed that such
system can describe different representations of $AdS_3$ algebra
such as massive, topologically massive, fractional spins and so on.
Recently one of us has shown the relation of such unfolded equations
with the Lagrangian formulation for massive bosonic fields
\cite{Zin15}. We begin with the frame-like gauge invariant
Lagrangian for the massive bosonic higher spin $s$ \cite{BSZ12a}
that includes a set of one-forms $\Omega^{\alpha(2k)}$, $1 \le k \le
s-1$ and zero-form $B^{\alpha(2)}$ (here and in what follows we use
a multispinor formalism, see below for notations and conventions).
Then it appears that to construct gauge invariant unfolded equations
one has to introduce a set of non gauge invariant zero forms
$B^{\alpha(2k)}$, $2 \le k \le s-1$ playing the roles of
Stueckelberg fields. At last the whole system is constructed by
addition of the infinite chain of gauge invariant zero-forms
$B^{\alpha(2k)}$, $k \ge s$ whose equations are in agreement with
the results of \cite{BPSS15}. Thus the complete  system contains the
following set of fields:
$$
\Omega^{\alpha(2k)}, B^{\alpha(2k)} \quad 1 \le k \le s-1,
\qquad B^{\alpha(2k)} \quad k \ge s
$$
Using our previous results on the frame-like gauge invariant
Lagrangian description of the massive spin $s+1/2$ fermionic field
\cite{BSZ14a} one can construct (see appendices C, D) an unfolded
formulation that also includes  a set of gauge and Stueckelberg fields
as well as infinite number of gauge invariant ones:
$$
\Phi^{\alpha(2k+1)}, \phi^{\alpha(2k+1)} \quad 0 \le k \le s-1, \qquad
\phi^{\alpha(2k+1)} \quad k \ge s
$$

Having in our disposal the unfolded equations for the massive bosonic
and fermionic fields we look for the supertransformations leaving
these equations invariant. Namely, following the example of the
massless scalar supermultiplet (1/2, 0) in $d=4$ \cite{PV10a} (see
also \cite{MV13}), we consider quadratic deformations of the unfolded
equations that have the form (schematically):
$$
0 = DB \oplus B \oplus F \Psi^\alpha, \qquad
0 = DF \oplus F \oplus B \Psi^\alpha
$$
where $\Psi^\alpha$ is the massless spin-3/2 gravitino. The
requirement that such deformations to be consistent determines all the
arbitrary coefficients. After that explicit expressions for the
supertransformations can be easily extracted and in the unfolded
formalism they have purely algebraic form:
$$
\delta B \sim F \zeta^\alpha, \qquad \delta F \sim B \zeta^\alpha
$$

The paper is organized as follows. For completeness and comparison in
Section 1 we consider two massless supermultiplets (s, s+1/2) and
(s, s-1/2) with arbitrary integer $s$. In-particular, these examples
clearly show specific properties of $d=3$ case related with the
factorization of $AdS_3$ superalgebra. We want to emphasize that the
supermultiplets (s, s+1/2) and (s, s-1/2) are essentially different.
In the first case the higher spin of the multiplet is half-integer,
while in the second case the higher spin is integer. Therefore these
two supermultiplets should be investigated separately. Section 2
contains a number of concrete (relatively) low spin examples of
massive supermultiplets. Here we consider (1, 1/2), (1/2, 0), (3/2, 1)
and (2, 3/2) ones. The main part of the paper contains sections 3 and
4 where we consider massive arbitrary spin supermultiplets (s, s+1/2)
and (s, s-1/2). The paper contains also four appendices devoted to the
unfolded equations that are necessary for the main part. Appendix A
gives unfolded equations for the $s=0,1/2,1,3/2,2$, Appendix B --- for
the massive spin-$s$ boson, while Appendices C and D --- for the
massive spins $s+1/2$ and $s-1/2$ fermions correspondingly.

\noindent
{\bf Notations and conventions.} We use a frame-like multispinor
formalism where all objects (one-forms or zero-forms) have local
indices which are completely symmetric spinor ones. To simplify
expressions we will use condensed notations for the spinor indices
such that e.g.
$$
\Omega^{\alpha(2k)} = \Omega^{(\alpha_1\alpha_2 \dots \alpha_{2k})}
$$
Also we will always assume that spinor indices denoted by the same
letter and placed on the same level are symmetrized, e.g.
$$
\Omega^{\alpha(2k)} \zeta^\alpha = \Omega^{(\alpha_1\dots \alpha_{2k}}
\zeta^{\alpha_{2k+1})}
$$
$AdS_3$ space will be described by the background frame (one-form)
$e^{\alpha(2)}$ and the covariant derivative $D$ normalized so that
$$
D \wedge D \zeta^\alpha = - \lambda^2 E^\alpha{}_\beta \zeta^\beta
$$
where two-form $E^{\alpha(2)}$ is defined as follows:
$$
e^{\alpha(2)} \wedge e^{\beta(2)} = \varepsilon^{\alpha\beta}
E^{\alpha\beta}
$$
In what follows the wedge product sign $\wedge$ will be omitted.

\section{Massless supermultiplets}

In this section we present an explicit construction for the
supermultiplets with massless higher spin fields.

\subsection{Kinematics}

The description of the massless bosonic spin-$s$ field in the
frame-like multispinor formalism requires a pair of one-forms
$\Omega^{\alpha(2s-2)}$ and $f^{\alpha(2s-2)}$. The free Lagrangian
(that is three-form in our formalism) describing such field living in
$AdS_3$ space has the form:
\begin{eqnarray}
{\cal L}_0 &=& (-1)^s [ (s-1) \Omega_{\alpha(2s-3)\beta}
e^\beta{}_\gamma \Omega^{\alpha(2s-3)\gamma} + \Omega_{\alpha(2s-2)} D
f^{\alpha(2s-2)} \nonumber \\
 && \qquad + \frac{(s-1)\lambda^2}{4} f_{\alpha(2s-3)\beta}
e^\beta{}_\gamma f^{\alpha(2s-3)\gamma} ]
\end{eqnarray}
This Lagrangian is invariant under the following local gauge
transformations:
\begin{eqnarray}
\delta \Omega^{\alpha(2s-2)} &=& D \eta^{\alpha(2s-2)} +
\frac{\lambda^2}{4} e^\alpha{}_\beta \xi^{\alpha(2s-3)\beta} \nonumber
\\
\delta f^{\alpha(2s-2)} &=& D \xi^{\alpha(2s-2)} + e^\alpha{}_\beta
\eta^{\alpha(2s-3)\beta}
\end{eqnarray}
where $\eta$ and $\xi$ are zero-forms completely symmetric in their
local indices.

Let us introduce new variables:
\begin{eqnarray}
\hat{\Omega}^{\alpha(2s-2)} &=& \Omega^{\alpha(2s-2)} +
\frac{\lambda}{2} f^{\alpha(2s-2)} \nonumber \\
\hat{f}^{\alpha(2s-2)} &=& \Omega^{\alpha(2s-2)} -
\frac{\lambda}{2} f^{\alpha(2s-2)}
\end{eqnarray}
and similarly for the parameters of the gauge transformations:
\begin{eqnarray}
\hat{\eta}^{\alpha(2s-2)} &=& \eta^{\alpha(2s-2)} + \frac{\lambda}{2}
\xi^{\alpha(2s-2)} \nonumber \\
\hat{\xi}^{\alpha(2s-2)} &=& \eta^{\alpha(2s-2)} - \frac{\lambda}{2}
\xi^{\alpha(2s-2)}
\end{eqnarray}
Then the Lagrangian can be rewritten as the sum of the two
independent parts:
\begin{eqnarray}
{\cal L}_0 &=& \frac{(-1)^s}{2\lambda} [ (s-1)\lambda
\hat{\Omega}_{\alpha(2s-3)\beta} e^\beta{}_\gamma
\hat{\Omega}^{\alpha(2s-3)\gamma} + \hat{\Omega}_{\alpha(2s-2)} D
\hat{\Omega}^{\alpha(2s-2)} \nonumber \\
 && \qquad + (s-1)\lambda \hat{f}_{\alpha(2s-3)\beta}
e^\beta{}_\gamma \hat{f}^{\alpha(2s-3)\gamma} -
\hat{f}_{\alpha(2s-2)} D \hat{f}^{\alpha(s-2)} ]
\end{eqnarray}
while the gauge transformations take the form:
\begin{eqnarray}
\delta \hat{\Omega}^{\alpha(2s-2)} &=& D \hat{\eta}^{\alpha(2s-2)} +
\frac{\lambda}{2} e^\alpha{}_\beta \hat{\eta}^{\alpha(2s-3)\beta}
\nonumber \\
\delta \hat{f}^{\alpha(2s-2)} &=& D \hat{\xi}^{\alpha(2s-2)} -
\frac{\lambda}{2} e^\alpha{}_\beta \hat{\xi}^{\alpha(2s-3)\beta}
\end{eqnarray}

Free Lagrangian for the massless fermionic field with spin $s$ has the
form:
\begin{equation}
{\cal L}_0 = \frac{i}{2} (-1)^{s-1/2} [ \Phi_{\alpha(2s-2)} D
\Phi^{\alpha(2s-2)} + (s-1)\lambda \Phi_{\alpha(2s-3)\beta}
e^\beta{}_\gamma \Phi^{\alpha(2s-3)\gamma}]
\end{equation}
It is invariant under the following local gauge transformations:
\begin{equation}
\delta \Phi^{\alpha(2s-2)} = D \xi^{\alpha(2s-2)} + \frac{\lambda}{2}
e^\alpha{}_\beta \xi^{\alpha(2s-3)\beta}
\end{equation}

\subsection{Supermultiplet ($s$, $s+1/2$)}

The free Lagrangian now is the sum of the free Lagrangians for the
bosonic and fermionic fields:
\begin{equation}
{\cal L}_0 = {\cal L}_0 (\Omega^{\alpha(2s-2)}, f^{\alpha(2s-2)}) +
{\cal L}_0 (\Phi^{\alpha(2s-1)})
\end{equation}
Let us consider the following ansatz for the global
supertransformations:
\begin{eqnarray}
\delta \Omega^{\alpha(2s-2)} &=& i\alpha_1 \Phi^{\alpha(2s-2)\beta}
\zeta_\beta \nonumber \\
\delta f^{\alpha(2s-2)} &=& i\alpha_2 \Phi^{\alpha(2s-2)\beta}
\zeta_\beta  \\
\delta \Phi^{\alpha(2s-1)} &=& \beta_1 \Omega^{\alpha(2s-2)}
\zeta^\alpha + \beta_2 f^{\alpha(2s-2)} \zeta^\alpha \nonumber
\end{eqnarray}
Recall that in $AdS_3$ space by global supertransformations we mean
the ones with the parameters satisfying the relation
\begin{equation}
D \zeta^\alpha = - \frac{\lambda}{2} e^\alpha{}_\beta \zeta^\beta
\end{equation}
Invariance of the Lagrangian requires:
$$
\alpha_1 = \frac{\lambda}{2} (2s-1)\beta_1, \qquad
\alpha_2 = (2s-1) \beta_1, \qquad
\beta_2 = \frac{\lambda}{2} \beta_1
$$
In terms of hatted variables it gives:
\begin{eqnarray}
\delta \hat{\Omega}^{\alpha(2s-2)} &=& i(2s-1)\lambda \beta_1
\Phi^{\alpha(2s-2)\beta} \zeta_\beta, \qquad
\delta \hat{f}^{\alpha(2s-2)} = 0 \nonumber \\
\delta \Phi^{\alpha(2s-1)} &=& \beta_1 \hat{\Omega}^{\alpha(2s-2)}
\zeta^\beta
\end{eqnarray}

\subsection{Supermultiplet ($s$, $s-1/2$)}

In this case the free Lagrangian has the form:
\begin{equation}
{\cal L}_0 = {\cal L}_0 (\Omega^{\alpha(2s-2)}, f^{\alpha(2s-2)}) +
{\cal L}_0 (\Phi^{\alpha(2s-3)})
\end{equation}
Let us consider the following ansatz for the supertransformations:
\begin{eqnarray}
\delta \Omega^{\alpha(2s-2)} &=& i\alpha_1 \Phi^{\alpha(2s-3)}
\zeta^\alpha \nonumber \\
\delta f^{\alpha(2s-2)} &=& i\alpha_2 \Phi^{\alpha(2s-3)} \zeta^\alpha
\\
\delta \Phi^{\alpha(2s-3)} &=& \beta_1 \Omega^{\alpha(2s-3)\beta}
\zeta_\beta + \beta_2 f^{\alpha(2s-3)\beta} \zeta_\beta \nonumber
\end{eqnarray}
Invariance of the Lagrangian requires:
$$
\alpha_1 = \frac{\lambda}{2} \alpha_2, \qquad
\beta_1 = 2(s-1)\alpha_2, \qquad
\beta_2 = 2(s-1)\frac{\lambda}{2}\alpha_2
$$
In terms of hatted variables it gives:
\begin{eqnarray}
\delta \hat{\Omega}^{\alpha(2s-2)} &=& i\lambda\alpha_2
\Phi^{\alpha(2s-3)} \zeta^\alpha, \qquad \delta \hat{f}^{\alpha(2s-2)}
= 0 \nonumber \\
\delta \Phi^{\alpha(2s-3)} &=& 2(s-1)\alpha_2
\hat{\Omega}^{\alpha(2s-3)\beta} \zeta_\beta
\end{eqnarray}

In both cases the results are consistent with the factorization of
superalgebra in $AdS_3$:
$$
{\cal A} \sim OSp(1,2) \otimes Sp(2)
$$

\section{Low spin examples}

In this section we consider examples of the massive low spins
supermultiplets with $0 \le s \le 2$. Unfolded equations for all these
fields are given in Appendix A.

\subsection{Supermultiplet (1, 1/2)}

The unfolded formulation of this supermultiplet requires bosonic
zero-forms $B^{\alpha(2k)}$, $k \ge 1$ as well as fermionic ones
$\phi^{\alpha(2k+1)}$, $k \ge 0$. As it was already explained in the
Introduction, our general procedure is to consider the deformation of
the initial unfolded equations corresponding to switching on a
background gravitino field $\Psi^\alpha$ satisfying the relation
$$
D \Psi^\alpha = - \frac{\lambda}{2} e^\alpha{}_\beta \Psi^\beta
$$
and to require that deformed equations remain to be consistent. This
in turn allows one easily extract the explicit form of the global
supertransformations leaving unfolded equations invariant.

Let us begin with the deformations for the bosonic equations:
\begin{eqnarray}
0 &=& D B^{\alpha(2k)} - e_{\beta(2)} B^{\alpha(2k)\beta(2)} - A_k
e^\alpha{}_\beta B^{\alpha(2k-1)\beta} - B_k e^{\alpha(2)}
B^{\alpha(2k-2)} \nonumber \\
 && - E_k \phi^{\alpha(2k)\beta} \Psi_\beta - F_k \phi^{\alpha(2k-1)}
\Psi^\alpha
\end{eqnarray}
Their consistency in the linear approximation (i.e. taking into
account quadratic terms in the consistency condition only) requires:
\begin{equation}
E_k = E_1, \qquad
F_k = - \frac{1}{2} [ (2k-1)C_{k-1} - 2(k-1)A_{k-1} -
\frac{\lambda}{2} ] E_1
\end{equation}
For the $A_k$ and $C_k$ corresponding to spin-1 ans spin-1/2 (see
appendix A) this gives
\begin{equation}
m = m_1 + \frac{\lambda}{2}, \qquad
F_k = - \frac{(k+1)}{2k(2k+1)} [ m_1 - k\lambda] E_1
\end{equation}

Similarly, deformations for the fermionic equations have the form:
\begin{eqnarray}
0 &=& D \phi^{\alpha(2k+1)} - e_{\beta(2)} \phi^{\alpha(2k+1)\beta(2)}
- C_k e^\alpha{}_\beta \phi^{\alpha(2k)\beta} - D_k e^{\alpha(2)}
\phi^{\alpha(2k-1)} \nonumber \\
 && - G_k B^{\alpha(2k+1)\beta} \Psi_\beta - H_k B^{\alpha(2k)}
\Psi^\alpha
\end{eqnarray}
Their consistency requires:
\begin{equation}
G_k = G_0, \qquad
H_k = - \frac{1}{2} [ 2kA_k - (2k-1)C_{k-1} - \frac{\lambda}{2} ] G_0
\end{equation}
and gives the same relation on masses. An explicit expression for
$H_k$ looks like:
\begin{equation}
H_k = \frac{k}{2(k+1)(2k+1)} [ m_1 + (k+1) \lambda ] G_0
\end{equation}

Thus we have found the supertransformations leaving unfolded equations
for massive spin-1 and spin-1/2 invariant:
\begin{eqnarray}
\delta B^{\alpha(2k)} &=& E_1 \phi^{\alpha(2k)\beta} \zeta_\beta + F_k
\phi^{\alpha(2k-1)} \zeta^\alpha \nonumber \\
\delta \phi^{\alpha(2k+1)} &=& G_0 B^{\alpha(2k+1)\beta} \zeta_\beta +
H_k B^{\alpha(2k)} \zeta^\alpha
\end{eqnarray}
In this, we have two arbitrary constants $E_1$ and $G_0$. Calculating
the commutator of these supertransformations one can fix $E_1G_0$ as a
normalization for the superalgebra. But to fix relative values for
$E_1$ and $G_0$ one has to construct appropriate Lagrangian formalism.

\subsection{Supermultiplet (1/2, 0)}

In this case we need the bosonic zero-forms $\pi^{\alpha(2k)}$ and the
fermionic ones $\phi^{\alpha(2k+1)}$, $k \ge 0$. Deformations for
their unfolded equations:
\begin{eqnarray}
0 &=& D \pi^{\alpha(2k)} - e_{\beta(2)} \pi^{\alpha(2k)\beta(2)} - B_k
e^{\alpha(2)} \pi^{\alpha(2k-2)} - E_k \phi^{\alpha(2k)\beta}
\Psi_\beta - F_k \phi^{\alpha(2k-1)} \Psi^\alpha \nonumber \\
0 &=& D \phi^{\alpha(2k+1)} - e_{\beta(2)} \phi^{\alpha(2k+1)\beta(2)}
- C_k e^\alpha{}_\beta \phi^{\alpha(2k)\beta} - D_k e^{\alpha(2)}
\phi^{\alpha(2k-1)} \\
 && - G_k \pi^{\alpha(2k+1)\beta} \Psi_\beta - H_k \pi^{\alpha(2k)}
\Psi^\alpha \nonumber
\end{eqnarray}
as well as all calculations are the same as in the previous case
except that now all $A_k=0$. We obtain:
\begin{eqnarray}
m_0{}^2 &=& m^2 - m\lambda - \frac{3}{4} \lambda^2 \nonumber \\
F_k &=& \frac{1}{2(2k+1)} [ m + (2k+1)\frac{\lambda}{2}] E_0 \\
H_k &=& - \frac{1}{2(2k+1)} [ m - (2k+1)\frac{\lambda}{2}] G_0
\nonumber
\end{eqnarray}

The supertransformations also have the same form:
\begin{eqnarray}
\delta \pi^{\alpha(2k)} &=& E_0 \phi^{\alpha(2k)\beta} \zeta_\beta +
F_k \phi^{\alpha(2k-1)} \zeta^\alpha \nonumber \\
\delta \phi^{\alpha(2k+1)} &=& G_0 \pi^{\alpha(2k+1)\beta} \zeta_\beta
+ H_k \pi^{\alpha(2k)} \zeta^\alpha
\end{eqnarray}

\subsection{Supermultiplet (3/2, 1)}

Unfolded formulation for the massive spin-3/2 requires one-form
$\Phi^\alpha$, Stueckelberg zero-form $\phi^\alpha$ as well as a
number of gauge invariant zero-forms $\phi^{\alpha(2k+1)}$, $k \ge 1$.
Let us consider the following deformations for the fermionic
equations:
\begin{eqnarray}
0 &=& D \Phi^\alpha + M e^\alpha{}_\beta \Phi^\beta + 2m
E^\alpha{}_\beta \phi^\beta - \alpha_1 e_{\beta(2)} B^{\beta(2)}
\Psi^\alpha \nonumber \\
0 &=& D \phi^\alpha + 2m \Phi^\alpha + M e^\alpha{}_\beta \phi^\beta -
e_{\beta(2)} \phi^{\alpha\beta(2)} - G_0 B^{\alpha\beta} \Psi_\beta \\
0 &=& D \phi^{\alpha(2k+1)} - e_{\beta(2)} \phi^{\alpha(2k+1)\beta(2)}
- C_k e^\alpha{}_\beta \phi^{\alpha(2k)\beta} - D_k e^{\alpha(2)}
\phi^{\alpha(2k-1)} \nonumber \\
 && - G_k B^{\alpha(2k+1)\beta} \Psi_\beta - H_k B^{\alpha(2k)}
\Psi^\alpha \nonumber
\end{eqnarray}
Consistency of the first equation requires:
\begin{equation}
\alpha_1 = \frac{mG_0}{(2M+\lambda)}, \qquad
M = m_1 - \frac{\lambda}{2}
\end{equation}
while the consistence of the remaining equations gives:
\begin{equation}
G_k = G_0, \qquad
H_k = - \frac{(k+2)}{2(k+1)(2k+1)} [ m_1 - (k+1)\lambda] G_0
\end{equation}

For the massive spin-1 unfolded equations and their deformations have
the same form as before:
\begin{eqnarray}
0 &=& D B^{\alpha(2k)} - e_{\beta(2)} B^{\alpha(2k)\beta(2)} - A_k
e^\alpha{}_\beta B^{\alpha(2k-1)\beta} - B_k e^{\alpha(2)}
B^{\alpha(2k-2)} \nonumber \\
 && - E_k \phi^{\alpha(2k)\beta} \Psi_\beta - F_k \phi^{\alpha(2k-1)}
\Psi^\alpha \end{eqnarray}
Their consistency gives the same mass relation and
\begin{equation}
E_k = E_1, \qquad
F_k = \frac{(k-1)}{2k(2k+1)} [ m_1 + k\lambda] E_1
\end{equation}

Thus, the supertransformations have the form ($k \ge 1$):
\begin{eqnarray}
\delta \Phi^\alpha &=& - \alpha_1 e_{\beta(2)} B^{\beta(2)}
\zeta^\alpha, \qquad \delta \phi^\alpha = G_0 B^{\alpha\beta}
\zeta_\beta \nonumber \\
\delta \phi^{\alpha(2k+1)} &=& G_0 B^{\alpha(2k+1)\beta} \zeta_\beta +
H_k B^{\alpha(2k)} \zeta^\alpha \\
\delta B^{\alpha(2k)} &=& E_1 \phi^{\alpha(2k)\beta} \zeta_\beta + F_k
\phi^{\alpha(2k-1)} \zeta^\alpha \nonumber
\end{eqnarray}
and also contain two arbitrary constants $E_1$ and $G_0$.

\subsection{Supermultiplet (2, 3/2)}

Unfolded formulation for the massive spin-2 requires one-form
$\Omega^{\alpha(2)}$, Stueckelberg zero-form $B^{\alpha(2)}$ and the
number of gauge invariant zero-forms $B^{\alpha(2k)}$, $k \ge 1$.
Deformations for the bosonic equations we take in the following form:
\begin{eqnarray}
0 &=& D \Omega^{\alpha(2)} + m_2 E^\alpha{}_\beta B^{\alpha\beta} +
\frac{M_2}{2} e^\alpha{}_\beta \Omega^{\alpha\beta} \nonumber \\
 && - \beta_1 \Phi^\alpha \Psi^\alpha - \beta_2 e^{\alpha(2)}
\phi_\beta \Psi^\beta \nonumber \\
0 &=& D B^{\alpha(2)} + m_2 \Omega^{\alpha(2)} + \frac{M_2}{2}
e^\alpha{}_\beta B^{\alpha\beta} - e_{\beta(2)} B^{\alpha(2)\beta(2)}
\nonumber \\
 && - E_1 \phi^{\alpha(2)\beta} \Psi_\beta - F_1 \phi^\alpha
\Psi^\alpha \\
0 &=& D B^{\alpha(2k)} - e_{\beta(2)} B^{\alpha(2k)\beta(2)} - A_k
e^\alpha{}_\beta B^{\alpha(2k-1)\beta} - B_k e^{\alpha(2)}
B^{\alpha(2k-2)} \nonumber \\
 && - E_k \phi^{\alpha(2k)\beta} \Psi_\beta - F_k \phi^{\alpha(2k-1)}
\Psi^\alpha \nonumber
\end{eqnarray}
The consistency of the first equation leads to
\begin{equation}
M_2 = M_1 - \frac{\lambda}{2}, \qquad
\beta_1 = \frac{m_1m_2}{(M_2+\lambda)}E_1, \qquad
\beta_2 = - \frac{m_2}{2} E_1
\end{equation}
while the consistency of the remaining equations produces
\begin{equation}
E_k = E_1, \qquad
F_k = - \frac{(k+2)}{2k(2k+1)} [ M_2 - k\lambda] E_1
\end{equation}

For the massive spin-3/2 we need the same set of fields as before
while the deformations for their equations look like:
\begin{eqnarray}
0 &=& D \Phi^\alpha + M_1 e^\alpha{}_\beta \Phi^\beta + 2m_1
E^\alpha{}_\beta \phi^\beta - \alpha_1 \Omega^{\alpha\beta} \Psi_\beta
- \alpha_2 e_{\beta(2)} B^{\beta(2)} \Psi^\alpha \nonumber \\
0 &=& D \phi^\alpha + 2m_1 \Phi^\alpha + M_1 e^\alpha{}_\beta
\phi^\beta - e_{\beta(2)} \phi^{\alpha\beta(2)} - G_0 B^{\alpha\beta}
\Psi_\beta \nonumber \\
0 &=& D \phi^{\alpha(2k+1)} - e_{\beta(2)} \phi^{\alpha(2k+1)\beta(2)}
- C_k e^\alpha{}_\beta \phi^{\alpha(2k)\beta} - D_k e^{\alpha(2)}
\phi^{\alpha(2k-1)} \\
 && - G_k B^{\alpha(2k+1)\beta} \Psi_\beta - H_k B^{\alpha(2k)}
\Psi^\alpha \nonumber
\end{eqnarray}
Consistency of the first one requires
\begin{equation}
\alpha_1 = - \frac{m_2}{2m_1}G_0, \qquad
\alpha_2 = \frac{m_1}{2(2M_1-\lambda)} G_0
\end{equation}
while the consistency of the remaining ones leads to
\begin{equation}
G_k = G_0, \qquad
H_k = \frac{(k-1)}{2(k+1)(2k+1)} [ M_2 + (k+1)\lambda] G_0
\end{equation}

The complete set of the supertransformations has the form:
\begin{eqnarray}
\delta \Omega^{\alpha(2)} &=& - \beta_1 \Phi^\alpha \zeta^\alpha -
\beta_2 e^{\alpha(2)} \phi_\beta \zeta^\beta \nonumber \\
\delta B^{\alpha(2k)} &=& E_1 \phi^{\alpha(2k)\beta} \zeta_\beta + F_k
\phi^{\alpha(2k-1)} \zeta^\alpha \nonumber \\
\delta \Phi^\alpha &=& - \alpha_1 \Omega^{\alpha\beta} \zeta_\beta -
\alpha_2 e_{\beta(2)} B^{\beta(2)} \zeta^\alpha \\
\delta \phi^{\alpha(2k+1)} &=& G_0 B^{\alpha(2k)\beta} \zeta_\beta +
H_k B^{\alpha(2k)} \zeta^\alpha \nonumber
\end{eqnarray}

\section{Supermultiplet ($s$, $s+1/2$)}

The unfolded formulation for the bosonic spin-$s$ field (Appendix B)
requires a set of one-forms $\Omega^{\alpha(2k)}$ and Stueckelberg
zero-forms $B^{\alpha(2k)}$, $1 \le k \le s-1$ as well as an infinite
number of gauge invariant zero-forms $B^{\alpha(2k)}$, $ k \ge s$.
Similarly, the unfolded formulation for the fermionic spin $s+1/2$
field (Appendix C) requires a set of one-forms $\Phi^{\alpha(2k+1)}$
and Stueckelberg zero-forms $\phi^{\alpha(2k+1)}$, $0 \le k \le s-1$
as well as an infinite number of gauge invariant zero-forms
$\phi^{\alpha(2k+1)}$, $k \ge s$.

Let us begin with the deformations of the equations for the gauge
invariant fermionic zero-forms ($k \ge s$):
\begin{eqnarray}
0 &=& D \phi^{\alpha(2k+1)} - e_{\beta(2)} \phi^{\alpha(2k+1)\beta(2)}
- C_k e^\alpha{}_\beta \phi^{\alpha(2k)\beta} - D_k e^{\alpha(2)}
\phi^{\alpha(2k-1)} \nonumber  \\
 && - G_k B^{\alpha(2k+1)\beta} \Psi_\beta - H_k B^{\alpha(2k)}
\Psi^\alpha
\end{eqnarray}
Their consistency gives the mass relation:
\begin{equation}
M_1 = M_2 - \frac{\lambda}{2}
\end{equation}
and fixes all coefficients in terms of one constant $G_s$:
\begin{equation}
G_k = G_s, \qquad
H_k = - \frac{(s+k+1)}{2(k+1)(2k+1)} [ M_2 - (k+1)\lambda] G_s, \qquad
k \ge s
\end{equation}
Note that the expression for $H_k$ is consistent with the particular
cases considered above.

Now we consider the deformations of the equations for the fermionic
one-forms:
\begin{eqnarray}
0 &=& D \Phi^{\alpha(2k+1)} + d_k e_{\beta(2)}
\Phi^{\alpha(2k+1)\beta(2)} + c_k e^\alpha{}_\beta
\Phi^{\alpha(2k)\beta} + \frac{d_{k-1}}{k(2k+1)} e^{\alpha(2)}
\Phi^{\alpha(2k-1)} \nonumber \\
&& + \alpha_k \Omega^{\alpha(2k+1)\beta} \Psi_\beta
+ \beta_k \Omega^{\alpha(2k)} \Psi^\alpha \\
0 &=& D \Phi^\alpha + d_0 e_{\beta(2)} \Phi^{\alpha\beta(2)} + c_0
e^\alpha{}_\beta \Phi^\beta + 4d_{(-1)}{}^2 E^\alpha{}_\beta
\phi^\beta + \alpha_0 \Omega^{\alpha\beta} \Psi_\beta
+ \beta_0 e_{\beta(2)} B^{\beta(2)} \Psi^\alpha  \nonumber
\end{eqnarray}
Their consistency requires (here
$\hat{\alpha}{}^2=2\hat{\beta}{}^2=(M_2-s\lambda)G_s{}^2$):
\begin{eqnarray}
\alpha_k{}^2 &=& \frac{(s-k-1)}{(2k+3)} [ M_2 + (k+1)\lambda ]
\hat{\alpha}^2 \nonumber \\
\beta_k{}^2 &=& \frac{(s+k+1)}{(k+1)(2k+1)^2} [ M_2 - (k+1)\lambda ]
\hat{\beta}^2 \\
\beta_0 &=& \frac{(s+1)}{2} [ M_2 - \lambda] \alpha_0 \nonumber
\end{eqnarray}

At last, the deformations of the equations for the Stueckelberg
fermionic zero-forms look like:
\begin{eqnarray}
0 &=& D \phi^{\alpha(2k+1)} + \Phi^{\alpha(2k+1)} + c_k
e^\alpha{}_\beta \phi^{\alpha(2k)\beta} + \frac{d_{k-1}}{k(2k+1)}
e^{\alpha(2)} \phi^{\alpha(2k-1)} + d_k e_{\beta(2)}
\phi^{\alpha(2k+1)\beta(2)} \nonumber \\
 && - G_k B^{\alpha(2k+1)\beta} \Psi_\beta - H_k B^{\alpha(2k)}
\Psi^\alpha \\
0 &=& D \phi^{\alpha(2s-1)} + \Phi^{\alpha(2s-1)} + c_{(s-1)}
e^\alpha{}_\beta \phi^{\alpha(2s-2)\beta} +
\frac{d_{(s-2)}}{(s-1)(2s-1)} e^{\alpha(2)} \phi^{\alpha(2s-3)}
\nonumber \\
 && - e_{\beta(2)} \phi^{\alpha(2s-1)\beta(2)}
 - G_{s-1} B^{\alpha(2s-1)\beta} \Psi_\beta - H_{s-1}
B^{\alpha(2s-2)} \Psi^\alpha \nonumber
\end{eqnarray}
Their consistency give
\begin{equation}
G_k = \alpha_k, \qquad H_k = \beta_k, \qquad k \le s-2
\end{equation}
$$
H_{s-1} = \beta_{s-1}, \qquad
G_s = G_{s-1} = \frac{(2s-1)\beta_{s-1}}{M_2-s\lambda}
$$

Let us now turn to the bosonic equations. Once again it is convenient
to begin with the deformations of the equations for the gauge
invariant zero-forms:
\begin{eqnarray}
0 &=& D B^{\alpha(2k)} - e_{\beta(2)} B^{\alpha(2k)\beta(2)} - A_k
e^\alpha{}_\beta B^{\alpha(2k-1)\beta} - B_k e^{\alpha(2)}
B^{\alpha(2k-2)} \nonumber \\
 && - E_k \phi^{\alpha(2k)\beta} \Psi_\beta - F_k \phi^{\alpha(2k-1)}
\Psi^\alpha
\end{eqnarray}
They give the same relation on masses and:
\begin{equation}
E_k = E_s, \qquad
F_k = - \frac{(s-k)}{2k(2k+1)} [ M_2 + k\lambda] E_s, \qquad
k \ge s
\end{equation}
Note that the expression for $F_k$ is also consistent with the
particular cases considered above.

Deformations of the equations for the bosonic one-forms look like:
\begin{eqnarray}
0 &=& D \Omega^{\alpha(2k)} + b_k e_{\beta(2)}
\Omega^{\alpha(2k)\beta(2)} + a_k e^\alpha{}_\beta
\Omega^{\alpha(2k-1)\beta} + \frac{b_{k-1}}{k(2k-1)} e^{\alpha(2)}
\Omega^{\alpha(2k-2)} \nonumber \\
 && + \gamma_k \Phi^{\alpha(2k)\beta} \Psi_\beta + \delta_k
\Phi^{\alpha(2k-1)} \Psi^\alpha \nonumber \\
0 &=& D \Omega^{\alpha(2)} + b_1 e_{\beta(2)}
\Omega^{\alpha(2)\beta(2)} + a_1 e^\alpha{}_\beta \Omega^{\alpha\beta}
+ 2b_0^2 E^\alpha{}_\beta B^{\alpha\beta} \\
 && + \gamma_1 \Phi^{\alpha(2)\beta} \Psi_\beta
+ \delta_1 \Phi^\alpha \Psi^\alpha + \gamma_0 e^{\alpha(2)}
\phi^\beta \Psi_\beta \nonumber
\end{eqnarray}
They give (here $\hat{\gamma}{}^2 =
2\hat{\delta}{}^2=E_s{}^2/2(M_2-s\lambda)$):
\begin{eqnarray}
\gamma_k{}^2 &=& \frac{(s+k+1)}{(k+1)} [ M_2 - (k+1)\lambda]
\hat{\gamma}^2 \nonumber \\
\delta_k{}^2 &=& \frac{(s-k)}{k^2(2k+1)} [ M_2 + k\lambda]
\hat{\delta}^2 \\
\gamma_0 &=& - (s+1)[M_2 - \lambda] \delta_1 \nonumber
\end{eqnarray}

At last, the deformations of the equations for the Stueckelberg
bosonic zero-forms:
\begin{eqnarray}
0 &=& D B^{\alpha(2k)} + \Omega^{\alpha(2k)} + a_k
e^\alpha{}_\beta B^{\alpha(2k-1)\beta} + \frac{b_{k-1}}{k(2k-1)}
e^{\alpha(2)} B^{\alpha(2k-2)} + b_k e_{\beta(2)}
B^{\alpha(2k)\beta(2)} \nonumber \\
 && - E_k \phi^{\alpha(2k)\beta} \Psi_\beta - F_k \phi^{\alpha(2k-1)}
\Psi^\alpha \nonumber \\
0 &=& D B^{\alpha(2s-2)} + \Omega^{\alpha(2s-2)} + a_{s-1}
e^\alpha{}_\beta B^{\alpha(2s-3)\beta} + \frac{b_{s-2}}{(s-1)(2s-3)}
e^{\alpha(2)} B^{\alpha(2s-4)} \\
 && - e_{\beta(2)} B^{\alpha(2s-2)\beta(2)}
 - E_{s-1} \phi^{\alpha(2s-2)\beta} \Psi_\beta - F_{s-1}
\phi^{\alpha(2s-3)} \Psi^\alpha  \nonumber
\end{eqnarray}
Their consistency requires:
\begin{equation}
E_k = \gamma_k, \qquad F_k = \delta_k, \qquad k \le s-1, \qquad
E_s = E_{s-1}
\end{equation}

Thus we have expressed all the coefficients in terms of just two
arbitrary constants (one can choose ($E_s,G_s$) or ($\hat{\alpha},
\hat{\gamma}$)). The complete set of supertransformations leaving all
unfolded equations invariant has the form:
\begin{eqnarray}\label{ST4}
\delta \Omega^{\alpha(2k)} &=& \gamma_k \Phi^{\alpha(2k)\beta}
\zeta_\beta + \delta_k \Phi^{\alpha(2k)} \zeta^\alpha \nonumber \\
\delta \Omega^{\alpha(2)} &=& \gamma_1 \Phi^{\alpha(2)\beta}
\zeta_\beta + \delta_1 \Phi^\alpha \zeta^\alpha + \gamma_0
e^{\alpha(2)} \phi^\beta \zeta_\beta \nonumber \\
\delta B^{\alpha(2k)} &=& E_k \phi^{\alpha(2k)\beta} \zeta_\beta + F_k
\phi^{\alpha(2k-1)} \zeta^\alpha \\
\delta \Phi^{\alpha(2k+1)} &=& \alpha_k \Omega^{\alpha(2k+1)\beta}
\zeta_\beta + \beta_k \Omega^{\alpha(2k)} \zeta^\alpha \nonumber \\
\delta \Phi^\alpha &=& \alpha_0 \Omega^{\alpha\beta} \zeta_\beta +
\beta_0 e_{\beta(2)} B^{\beta(2)} \zeta^\alpha \nonumber \\
\delta \phi^{\alpha(2k+1)} &=& G_k B^{\alpha(2k+1)\beta} \zeta_\beta +
H_k B^{\alpha(2k)} \zeta^\alpha \nonumber
\end{eqnarray}

\section{Supermultiplet ($s$, $s-1/2$)}

In this case the bosonic equations are the same as before while the
unfolded formulation for the fermionic spin $s-1/2$ field (Appendix D)
requires a set of one-forms $\Phi^{\alpha(2k+1)}$ and Stueckelberg
zero-forms $\phi^{\alpha(2k+1)}$, $0 \le k \le s-2$ as well as an
infinite number of gauge invariant zero-forms $\phi^{\alpha(2k+1)}$,
$k \ge s-1$.

Deformations for gauge invariant fermionic zero-forms:
\begin{eqnarray}
0 &=& D \phi^{\alpha(2k+1)} - e_{\beta(2)} \phi^{\alpha(2k+1)\beta(2)}
- C_k e^\alpha{}_\beta \phi^{\alpha(2k)\beta} - D_k e^{\alpha(2)}
\phi^{\alpha(2k-1)} \nonumber  \\
 && - G_k B^{\alpha(2k+1)\beta} \Psi_\beta - H_k B^{\alpha(2k)}
\Psi^\alpha
\end{eqnarray}
Their consistency gives a mass relation:
\begin{equation}
M_1 = M_2 + \frac{\lambda}{2}
\end{equation}
as well as fixes all coefficients in terms of $G_{s-1}$:
\begin{equation}
G_k = G_{s-1}, \qquad
H_k = - \frac{(s-k-1)}{2(k+1)(2k+1)} [ M_2 + (k+1)\lambda] G_{s-1}
\end{equation}

Deformations for the fermionic one-forms look like:
\begin{eqnarray}
0 &=& D \Phi^{\alpha(2k+1)} + d_k e_{\beta(2)}
\Phi^{\alpha(2k+1)\beta(2)} + c_k e^\alpha{}_\beta
\Phi^{\alpha(2k)\beta} + \frac{d_{k-1}}{k(2k+1)} e^{\alpha(2)}
\Phi^{\alpha(2k-1)} \nonumber \\
&& + \alpha_k \Omega^{\alpha(2k+1)\beta} \Psi_\beta
+ \beta_k \Omega^{\alpha(2k)} \Psi^\alpha \\
0 &=& D \Phi^\alpha + d_0 e_{\beta(2)} \Phi^{\alpha\beta(2)} + c_0
e^\alpha{}_\beta \Phi^\beta + 4d_{(s-1)}{}^2 E^\alpha{}_\beta
\phi^\beta + \alpha_0 \Omega^{\alpha\beta} \Psi_\beta
+ \beta_0 e_{\beta(2)} B^{\beta(2)} \Psi^\alpha  \nonumber
\end{eqnarray}
Their consistency requires (here
$\hat{\alpha}{}^2=2\hat{\beta}{}^2=G_{s-1}{}^2/(M_2-(s-1)\lambda)$):
\begin{eqnarray}
\alpha_k{}^2 &=& \frac{(s+k+1)}{(2k+3)} [ M_2 - (k+1)\lambda]
\hat{\alpha}^2 \nonumber \\
\beta_k{}^2 &=& \frac{(s-k-1)}{(k+1)(2k+1)^2} [M_2 + (k+1)\lambda]
\hat{\beta}^2 \\
\beta_0 &=& \frac{(s-1)}{2} [ M_2 + \lambda] \alpha_0 \nonumber
\end{eqnarray}

At the same time deformations for the Stueckelberg fermionic
zero-forms:
\begin{eqnarray}
0 &=& D \phi^{\alpha(2k+1)} + \Phi^{\alpha(2k+1)} + c_k
e^\alpha{}_\beta \phi^{\alpha(2k)\beta} + \frac{d_{k-1}}{k(2k+1)}
e^{\alpha(2)} \phi^{\alpha(2k-1)} + d_k e_{\beta(2)}
\phi^{\alpha(2k+1)\beta(2)} \nonumber \\
 && - G_k B^{\alpha(2k+1)\beta} \Psi_\beta - H_k B^{\alpha(2k)}
\Psi^\alpha \\
0 &=& D \phi^{\alpha(2s-3)} + \Phi^{\alpha(2s-3)} + c_{(s-2)}
e^\alpha{}_\beta \phi^{\alpha(2s-4)\beta} +
\frac{d_{(s-3)}}{(s-2)(2s-3)} e^{\alpha(2)} \phi^{\alpha(2s-5)}
\nonumber \\
 && - e_{\beta(2)} \phi^{\alpha(2s-3)\beta(2)}
 - G_{s-2} B^{\alpha(2s-3)\beta} \Psi_\beta - H_{s-2}
B^{\alpha(2s-4)} \Psi^\alpha  \nonumber
\end{eqnarray}
They give:
\begin{equation}
G_k = \alpha_k, \qquad
H_k = \beta_k, \quad k \le s-2, \qquad
G_{s-1} = G_{s-2}
\end{equation}

Deformations for bosonic gauge invariant zero-forms:
\begin{eqnarray}
0 &=& D B^{\alpha(2k)} - e_{\beta(2)} B^{\alpha(2k)\beta(2)} - A_k
e^\alpha{}_\beta B^{\alpha(2k-1)\beta} - B_k e^{\alpha(2)}
B^{\alpha(2k-2)} \nonumber \\
 && - E_k \phi^{\alpha(2k)\beta} \Psi_\beta - F_k \phi^{\alpha(2k-1)}
\Psi^\alpha
\end{eqnarray}
Their consistency gives the same mass relation as before and leads to
the following expressions for all coefficients in terms of $E_s$:
\begin{equation}
E_k = E_s, \qquad
F_k = - \frac{(s+k)}{2k(2k+1)} [ M_2 - k\lambda] E_s, \qquad
k \ge s
\end{equation}

Deformations for bosonic one-forms:
\begin{eqnarray}
0 &=& D \Omega^{\alpha(2k)} + b_k e_{\beta(2)}
\Omega^{\alpha(2k)\beta(2)} + a_k e^\alpha{}_\beta
\Omega^{\alpha(2k-1)\beta} + \frac{b_{k-1}}{k(2k-1)} e^{\alpha(2)}
\Omega^{\alpha(2k-2)} \nonumber \\
 && + \gamma_k \Phi^{\alpha(2k)\beta} \Psi_\beta + \delta_k
\Phi^{\alpha(2k-1)} \Psi^\alpha, \qquad \gamma_{s-1} = 0 \nonumber \\
0 &=& D \Omega^{\alpha(2)} + b_1 e_{\beta(2)}
\Omega^{\alpha(2)\beta(2)} + a_1 e^\alpha{}_\beta \Omega^{\alpha\beta}
+ 2b_0^2 E^\alpha{}_\beta B^{\alpha\beta} \\
 && + \gamma_1 \Phi^{\alpha(2)\beta} \Psi_\beta
+ \delta_1 \Phi^\alpha \Psi^\alpha + \gamma_0 e^{\alpha(2)}
\phi^\beta \Psi_\beta \nonumber
\end{eqnarray}
Their consistence requires (here
$\hat{\gamma}{}^2=2\hat{\delta}{}^2=\frac12(M_2-(s-1)\lambda)E_s^2$):
\begin{eqnarray}
\gamma_k{}^2 &=& \frac{(s-k-1)}{(k+1)} [ M_2 + (k+1)\lambda]
\hat{\gamma}^2 \nonumber \\
\delta_k{}^2 &=& \frac{(s+k)}{k^2(2k+1)} [ M_2 - k\lambda]
\hat{\delta}^2 \\
\gamma_0 &=& - (s-1) [ M_2 + \lambda] \delta_1 \nonumber
\end{eqnarray}

At last deformations for the Stueckelberg bosonic zero-forms:
\begin{eqnarray}
0 &=& D B^{\alpha(2k)} + \Omega^{\alpha(2k)} + a_k
e^\alpha{}_\beta B^{\alpha(2k-1)\beta} + \frac{b_{k-1}}{k(2k-1)}
e^{\alpha(2)} B^{\alpha(2k-2)} + b_k e_{\beta(2)}
B^{\alpha(2k)\beta(2)} \nonumber \\
 && - E_k \phi^{\alpha(2k)\beta} \Psi_\beta - F_k \phi^{\alpha(2k-1)}
\Psi^\alpha \nonumber \\
0 &=& D B^{\alpha(2s-2)} + \Omega^{\alpha(2s-2)} + a_{s-1}
e^\alpha{}_\beta B^{\alpha(2s-3)\beta} + \frac{b_{s-2}}{(s-1)(2s-3)}
e^{\alpha(2)} B^{\alpha(2s-4)} \\
 && - e_{\beta(2)} B^{\alpha(2s-2)\beta(2)}  - E_{s-1}
\phi^{\alpha(2s-2)\beta} \Psi_\beta - F_{s-1}
\phi^{\alpha(2s-3)} \Psi^\alpha \nonumber
\end{eqnarray}
We obtain:
\begin{equation}
E_k = \gamma_k, \qquad F_k = \delta_k, \quad k \le s-2, \qquad
F_{s-1} = \delta_{s-1}
\end{equation}
$$
E_s = E_{s-1} = \frac{\gamma_{s-1}}{b_{s-2}} =
\frac{2(s-1)\delta_{s-1}}{(M_2-(s-1)\lambda)}
$$

Supertransformations look the same as before (\ref{ST4}) (taking
into account that $\Phi^{\alpha(2s-1)}$ is absent now), but with the
new expressions for all coefficients.

\section{Summary and Conclusion}

In this paper we have presented the systematic derivation of the
unfolded equations for three dimensional supersymmetric higher spin
theory. Final results are formulated in sections 3 and 4. In
particular, the equations (\ref{ST4}) solves a problem of the
supersymmetry transformations for the supermultiplets $(s, s+1/2)$
and $(s, s-1/2)$ respectively. Thus, we have the complete system of
the unfolded equations and the corresponding supersymmetry
transformations for the most general massive $(1,0)$ supersymmetric
field model in $AdS_3$ space.

As it is typical for the unfolded formalism, our results contain an
infinite number of fields and so an infinite number of equations which
are non-Lagrangian ones. As we pointed out in the Introduction, the
same problem can be analysed on the base of approach developed in the
papers \cite{BSZ15}, \cite{BSZ12a}, \cite{BSZ14a}. Although this
approach looks like more complicated, it operates with a finite number
of fields and there are no reasons to expect that the final equations
of motion should be non-Lagrangian. Therefore we suppose that the
unfolded equations, obtained here, can be somehow reformulated,
perhaps with eliminating some auxiliary fields, so that we will obtain
the Lagrangian formulation. We guess that this aspect deserves a
special consideration.

The equations and the supersymmetry transformations are obtained here
in the component formulation. Both the equations and the supersymmetry
transformations include all the fields which are necessary for
consistent supersymmetric higher spin dynamics. However, they do not
contain the auxiliary fields needed for off-shell supersymmetry. We
suppose that a solution of off-shell supersymmetry problem can be
realized on the base of appropriate superfield formalism. Developing
such a formalism is one of the open problem in the three dimensional
massive supersymmetric higher spin theory.

After the first version of this paper has been appeared in ArXiv, we
were informed by S.M. Kuzenko that he and M. Tsulaia have
constructed the massive higher spin off-shell ${\cal N}=1$
supermultiplets on $AdS_3$.

\section*{Acknowledgments}
The authors acknowledge S.M. Kuzenko for useful comments. I.L.B and
T.V.S are grateful to the RFBR grant, project No. 15-02-03594-a for
partial support. Their research was also supported in parts by
Russian Ministry of Education and Science, project 2014/387.122.
I.L.B and Yu.M.Z are thankful to MIAPP program "Higher-Spin Theory
and Duality" where this paper was finalized. T.V.S acknowledges
partial support from the President of Russia grant for young
scientists No. MK-6453.2015.2. Work of Yu.M.Z was supported in parts
by RFBR grant, project No. 14-02-01172.

\section{Appendix A. Unfolded equations for low spins}

In this appendix we have collected all unfolded equations for the
bosonic and fermionic fields with spin $0 \le s \le 2$.

\subsection{Spin 0}

Unfolded formulation for the spin-0 is very well known. It requires an
infinite set of zero-forms $\pi^{\alpha(2k)}$, $k \ge 0$ satisfying
the equations:
\begin{equation}
0 = D \pi^{\alpha(2k)} - e_{\beta(2)} \pi^{\alpha(2k)\beta(2)} - B_k
e^{\alpha(2)} \pi^{\alpha(2k-2)}
\end{equation}
Their consistency requires:
\begin{equation}
(2k+3) B_{k+1} = (2k-1)B_k + \frac{\lambda^2}{2}
\end{equation}
and leads to the solution:
\begin{equation}
B_k = - \frac{m_0{}^2 - (k^2-1)\lambda^2}{2(4k^2-1)}
\end{equation}

\subsection{Spin 1/2}

Similarly, the unfolded formulation for the spin-1/2 requires an
infinite umber of fermionic zero-forms $\Phi^{\alpha(2k+1)}$, $k \ge
0$
satisfying the equations:
\begin{equation}
0 = D \phi^{\alpha(2k+1)} - e_{\beta(2)} \phi^{\alpha(2k+1)\beta(2)}
- C_k e^\alpha{}_\beta \phi^{\alpha(2k)\beta} - D_k e^{\alpha(2)}
\phi^{\alpha(2k-1)}
\end{equation}
Their consistency requires:
\begin{equation}
(2k+5)C_{k+1} = (2k+1)C_k, \qquad
(k+2)D_{k+1} = kD_k - C_k{}^2 + \frac{\lambda^2}{4}
\end{equation}
and leads to the solution:
\begin{equation}
C_k = - \frac{m}{(2k+3)(2k+1)}, \qquad
D_k = - \frac{m^2}{2(2k+1)^2} + \frac{\lambda^2}{8}
\end{equation}

\subsection{Spin 1}

In this case one also needs an infinite number of bosonic zero-forms
$B^{\alpha(2k)}$ but now with $k \ge 1$. The unfolded equations have
the form:
\begin{equation}
0 = D B^{\alpha(2k)} - e_{\beta(2)} B^{\alpha(2k)\beta(2)} - A_k
e^\alpha{}_\beta B^{\alpha(2k-1)\beta} - B_k e^{\alpha(2)}
B^{\alpha(2k-2)}
\end{equation}
Their consistency implies the relations on the coefficients $A$, $B$:
\begin{equation}
(k+2)A_{k+1} = kA_k, \qquad
(2k+3)B_{k+1} = (2k-1)B_k - 2A_k{}^2 + \frac{\lambda^2}{2}
\end{equation}
which have the following solution:
\begin{equation}
A_k = - \frac{m_1}{2k(k+1)}, \qquad
B_k = - \frac{(k^2-1)}{2(4k^2-1)} [ \frac{m_1{}^2}{k^2} - \lambda^2]
\end{equation}

\subsection{Spin 3/2}

The unfolded formulation for massive spin-3/2 already has the general
pattern. Namely, it requires one-form $\Phi^\alpha$ and Stueckelberg
zero-form $\phi^\alpha$ as well as an infinite number of gauge
invariant zero-forms $\phi^{\alpha(2k+1)}$, $k \ge 1$. The unfolded
equations for the first two fields look like:
\begin{eqnarray}
0 &=& D \Phi^\alpha + M e^\alpha{}_\beta \Psi^\beta + 2m
E^\alpha{}_\beta \phi^\beta \nonumber \\
0 &=& D \phi^\alpha + 2m \Phi^\alpha + M e^\alpha{}_\beta \phi^\beta -
e_{\beta(2)} \phi^{\alpha\beta(2)}
\end{eqnarray}
where:
$$
M^2 = m^2 + \frac{\lambda^2}{4}
$$
Equations for the remaining fields have the same form as in the
spin-1/2 case:
\begin{equation}
0 = D \phi^{\alpha(2k+1)} - e_{\beta(2)} \phi^{\alpha(2k+1)\beta(2)} -
C_k e^\alpha{}_\beta \phi^{\alpha(2k)\beta} - D_k e^{\alpha(2)}
\phi^{\alpha(2k-1)}
\end{equation}
Consistency conditions are also the same as for the spin-1/2 case but
their solution now:
\begin{equation}
C_k = - \frac{3M}{(2k+3)(2k+1)}, \qquad
D_k = - \frac{(k+2)(k-1)}{8k(k+1)} [ \frac{4M^2}{(2k+1)^2} -
\lambda^2]
\end{equation}

\subsection{Spin 2}

In this case we also need the one-form $\Omega^{\alpha(2)}$ and
Stueckelberg zero-form $B^{\alpha(2)}$ as well as an infinite number
of gauge invariant zero-forms $B^{\alpha(2k)}$, $k \ge 2$. The
equations for the first two fields:
\begin{eqnarray}
0 &=& D \Omega^{\alpha(2)} + m E^\alpha{}_\beta B^{\alpha\beta} +
\frac{M}{2} e^\alpha{}_\beta \Omega^{\alpha\beta} \nonumber \\
0&=& D B^{\alpha(2)} + m \Omega^{\alpha(2)} + \frac{M}{2}
e^\alpha{}_\beta B^{\alpha\beta} - e_{\beta(2)} B^{\alpha(2)\beta(2)}
\end{eqnarray}
where:
$$
M^2 = m^2 + \lambda^2
$$
The equations for the remaining fields have the same form as in the
spin-1 case:
\begin{equation}
0 = D B^{\alpha(2k)} - e_{\beta(2)} B^{\alpha(2k)\beta(2)} - A_k
e^\alpha{}_\beta B^{\alpha(2k-1)\beta} - B_k e^{\alpha(2)}
B^{\alpha(2k-2)}
\end{equation}
Consistency conditions are also the same as for spin-1, but their
solution now:
\begin{equation}
A_k = - \frac{M}{k(k+1)}, \qquad
B_k = - \frac{(k^2-4)}{2(4k^2-1)} [ \frac{M^2}{k^2} - \lambda^2 ]
\end{equation}

\section{Appendix B. Unfolded equations for the bosonic spin-$s$
field}

The unfolded formulation for the bosonic spin-$s$ field requires a set
of one-forms $\Omega^{\alpha(2k)}$ and Stueckelberg zero-forms
$B^{\alpha(2k)}$, $1 \le k \le s-1$ as well as an infinite number of
gauge invariant zero-forms $B^{\alpha(2k)}$, $ k \ge s$.

Equations for one-forms ($2 \le k \le s-1$ , $b_{s-1} = 0$):
\begin{eqnarray}
0 &=& D \Omega^{\alpha(2k)} + b_k e_{\beta(2)}
\Omega^{\alpha(2k)\beta(2)} + a_k e^\alpha{}_\beta
\Omega^{\alpha(2k-1)\beta} + \frac{b_{k-1}}{k(2k-1)} e^{\alpha(2)}
\Omega^{\alpha(2k-2)} \nonumber \\
0 &=& D \Omega^{\alpha(2)} + b_1 e_{\beta(2)}
\Omega^{\alpha(2)\beta(2)} + a_1 e^\alpha{}_\beta \Omega^{\alpha\beta}
+ 2b_0^2 E^\alpha{}_\beta B^{\alpha\beta}
\end{eqnarray}
Here
$$
M_2{}^2 = m_2{}^2 + (s-1)^2 \lambda^2
$$
\begin{equation}
a_k = \frac{M_2s}{2k(k+1)}, \qquad
b_k{}^2 = \frac{(s-k-1)(s+k+1)}{2(k+1)(2k+3)} [ M_2{}^2 - (k+1)^2
\lambda^2]
\end{equation}
Equations for the Stueckelberg zero-forms ($1 \le k \le s-2$):
\begin{eqnarray}
0 &=& D B^{\alpha(2k)} + \Omega^{\alpha(2k)} + a_k
e^\alpha{}_\beta B^{\alpha(2k-1)\beta} + \frac{b_{k-1}}{k(2k-1)}
e^{\alpha(2)} B^{\alpha(2k-2)} + b_k e_{\beta(2)}
B^{\alpha(2k)\beta(2)} \nonumber \\
0 &=& D B^{\alpha(2s-2)} + \Omega^{\alpha(2s-2)} + a_{s-1}
e^\alpha{}_\beta B^{\alpha(2s-3)\beta} + \frac{b_{s-2}}{(s-1)(2s-3)}
e^{\alpha(2)} B^{\alpha(2s-4)} \\
 && - e_{\beta(2)} B^{\alpha(2s-2)\beta(2)} \nonumber
\end{eqnarray}
Equations for gauge invariant zero-forms:
\begin{equation}
0 = D B^{\alpha(2k)} - e_{\beta(2)} B^{\alpha(2k)\beta(2)} - A_k
e^\alpha{}_\beta B^{\alpha(2k-1)\beta} - B_k e^{\alpha(2)}
B^{\alpha(2k-2)}
\end{equation}
Here
\begin{equation}
A_k = - \frac{Ms}{2k(k+1)}, \qquad
B_k = - \frac{(k^2-s^2)}{2(4k^2-1)} [ \frac{M^2}{k^2} - \lambda^2 ]
\end{equation}

\section{Appendix C. Unfolded equations for the fermionic spin
($s+1/2$) field}

The unfolded formulation for the fermionic spin $s+1/2$
field requires a set of one-forms $\Phi^{\alpha(2k+1)}$ and
Stueckelberg zero-forms $\phi^{\alpha(2k+1)}$, $0 \le k \le s-1$
as well as an infinite number of gauge invariant zero-forms
$\phi^{\alpha(2k+1)}$, $k \ge s$.

Equations for one-forms ($1 \le k \le s-1$, $d_{s-1} = 0$):
\begin{eqnarray}
0 &=& D \Phi^{\alpha(2k+1)} + d_k e_{\beta(2)}
\Phi^{\alpha(2k+1)\beta(2)} + c_k e^\alpha{}_\beta
\Phi^{\alpha(2k)\beta} + \frac{d_{k-1}}{k(2k+1)} e^{\alpha(2)}
\Phi^{\alpha(2k-1)} \nonumber \\
0 &=& D \Phi^\alpha + d_0 e_{\beta(2)} \Phi^{\alpha\beta(2)} + c_0
e^\alpha{}_\beta \Phi^\beta + 4d_{(-1)}{}^2 E^\alpha{}_\beta
\phi^\beta
\end{eqnarray}
Here
$$
M_1{}^2 = m_1{}^2 + (s-\frac{1}{2})^2 \lambda^2
$$
\begin{equation}
c_k = \frac{(2s+1)M_1}{(2k+1)(2k+3)}, \qquad
d_k{}^2 = \frac{(s-k-1)(s+k+2)}{2(2k+3)(k+2)}
[ M_1{}^2 - \frac{(2k+3)^2}{4} \lambda^2]
\end{equation}
Equations for the Stueckelberg zero-forms ($0 \le k \le s-2$):
\begin{eqnarray}
0 &=& D \phi^{\alpha(2k+1)} + \Phi^{\alpha(2k+1)} + c_k
e^\alpha{}_\beta \phi^{\alpha(2k)\beta} + \frac{d_{k-1}}{k(2k+1)}
e^{\alpha(2)} \phi^{\alpha(2k-1)} + d_k e_{\beta(2)}
\phi^{\alpha(2k+1)\beta(2)} \nonumber \\
0 &=& D \phi^{\alpha(2s-1)} + \Phi^{\alpha(2s-1)} + c_{(s-1)}
e^\alpha{}_\beta \phi^{\alpha(2s-2)\beta} +
\frac{d_{(s-2)}}{(s-1)(2s-1)} e^{\alpha(2)} \phi^{\alpha(2s-3)} \\
 && - e_{\beta(2)} \phi^{\alpha(2s-1)\beta(2)} \nonumber
\end{eqnarray}
Equations for gauge invariant zero-forms:
\begin{equation}
0 = D \phi^{\alpha(2k+1)} - e_{\beta(2)} \phi^{\alpha(2k+1)\beta(2)} -
C_k e^\alpha{}_\beta \phi^{\alpha(2k)\beta} - D_k e^{\alpha(2)}
\phi^{\alpha(2k-1)}
\end{equation}
Here
\begin{equation}
C_k = - \frac{(2s+1)M_1}{(2k+1)(2k+3)}, \qquad
D_k = - \frac{(k-s)(k+s+1)}{2k(k+1)} [ \frac{M_1{}^2}{(2k+1)^2} -
\frac{\lambda^2}{4} ]
\end{equation}

\section{Appendix D. Unfolded equations for the fermionic spin
($s-1/2$) field}

The unfolded formulation for the fermionic spin $s-1/2$ field requires
a set of one-forms $\Phi^{\alpha(2k+1)}$ and Stueckelberg zero-forms
$\phi^{\alpha(2k+1)}$, $0 \le k \le s-2$ as well as an infinite number
of gauge invariant zero-forms $\phi^{\alpha(2k+1)}$, $k \ge s-1$.

Equations for the one-forms ($1 \le k \le s-2$, $d_{s-2} = 0$):
\begin{eqnarray}
0 &=& D \Psi^{\alpha(2k+1)} + d_k e_{\beta(2)}
\Phi^{\alpha(2k+1)\beta(2)} + c_k e^\alpha{}_\beta
\Phi^{\alpha(2k)\beta} + \frac{d_{k-1}}{k(2k+1)} e^{\alpha(2)}
\Phi^{\alpha(2k-1)}  \nonumber \\
0 &=& D \Phi^\alpha + d_0 e_{\beta(2)} \Phi^{\alpha\beta(2)} + c_0
e^\alpha{}_\beta \Phi^\beta + 4d_{(-1)}{}^2 E^\alpha{}_\beta
\phi^\beta
\end{eqnarray}
Here
$$
M_1{}^2 = m_1{}^2 + (s-\frac{3}{2})^2 \lambda^2
$$
\begin{equation}
c_k = \frac{(2s-1)M_1}{(2k+1)(2k+3)}, \qquad
d_k{}^2 = \frac{(s-k-2)(s+k+1)}{2(2k+3)(k+2)}
[ M_1{}^2 - \frac{(2k+3)^2}{4} \lambda^2]
\end{equation}
Equations for the Stueckelberg zero-forms ($0 \le k \le s-3$):
\begin{eqnarray}
0 &=& D \phi^{\alpha(2k+1)} + \Phi^{\alpha(2k+1)} + c_k
e^\alpha{}_\beta \phi^{\alpha(2k)\beta} + \frac{d_{k-1}}{k(2k+1)}
e^{\alpha(2)} \phi^{\alpha(2k-1)} + d_k e_{\beta(2)}
\phi^{\alpha(2k+1)\beta(2)} \nonumber \\
0 &=& D \phi^{\alpha(2s-3)} + \Phi^{\alpha(2s-3)} + c_{(s-2)}
e^\alpha{}_\beta \phi^{\alpha(2s-4)\beta} +
\frac{d_{(s-3)}}{(s-2)(2s-3)} e^{\alpha(2)} \phi^{\alpha(2s-5)} \\
 && - e_{\beta(2)} \phi^{\alpha(2s-3)\beta(2)} \nonumber
\end{eqnarray}
Equations for gauge invariant zero-forms:
\begin{equation}
0 = D \phi^{\alpha(2k+1)} - e_{\beta(2)} \phi^{\alpha(2k+1)\beta(2)} -
C_k e^\alpha{}_\beta \phi^{\alpha(2k)\beta} - D_k e^{\alpha(2)}
\phi^{\alpha(2k-1)}
\end{equation}
Here
\begin{equation}
C_k = - \frac{(2s-1)M_1}{(2k+1)(2k+3)}, \qquad
D_k = - \frac{(k-s+1)(k+s)}{2k(k+1)} [ \frac{M_1{}^2}{(2k+1)^2} -
\frac{\lambda^2}{4} ]
\end{equation}

\end{document}